\documentclass[10pt]{iopart}

\expandafter\let\csname equation*\endcsname\relax
\expandafter\let\csname endequation*\endcsname\relax

\usepackage{textcomp}
\usepackage{graphicx}
\usepackage[abs]{overpic}
\usepackage{float}
\usepackage{lscape}
\usepackage[caption=false]{subfig}
\usepackage[version=4]{mhchem}
\PassOptionsToPackage{hyphens}{url}
\usepackage{hyperref}
\usepackage{xcolor}
\usepackage{textgreek}
\usepackage{booktabs}
\usepackage[normalem]{ulem}
\usepackage{layouts}
\usepackage{xtab}
\usepackage{changepage}

\begin{document}
\begin{adjustwidth}{1.275cm}{-1.275cm}
\title[Decoupling NO production and UV emission intensity over the E-H mode transition in a low-pressure ICP]{Decoupling NO production and UV emission intensity over the E-H mode transition in a low-pressure inductively coupled plasma device}

\author{Lars Schücke\textsuperscript{1,2}, Angie Natalia Torres Segura\textsuperscript{1}, Ihor Korolov\textsuperscript{1}, Peter Awakowicz\textsuperscript{1} and Andrew R. Gibson\textsuperscript{2,3}}

\address{\textsuperscript{1}Chair of Applied Electrodynamics and Plasma Technology, Faculty of Electrical Engineering and Information Technology, Ruhr University Bochum, Germany}
\address{\textsuperscript{2}Research Group for Biomedical Plasma Technology, Faculty of Electrical Engineering and Information Technology, Ruhr University Bochum, Germany}
\address{\textsuperscript{3}York Plasma Institute, School of Physics, Engineering and Technology, University of York, UK}

\ead{schuecke@aept.rub.de}

\ead{andrew.gibson@york.ac.uk}

\vspace{10pt}
\begin{indented}
\item[]June 2024
\end{indented}

\begin{abstract}
A low-pressure double-inductively coupled plasma device is used to study the fundamental plasma parameters, plasma chemistry, and UV photon emission from the first excited state of nitric oxide, NO(A), in gas mixtures of nitrogen and oxygen. In addition to the gas mixture, rf power and gas pressure are varied, and the E-H mode transition of the inductively coupled plasma is studied specifically. The gas temperature and UV photon emission are measured by optical emission spectroscopy (OES), the absolute density of the nitric oxide electronic ground state by laser-induced fluorescence (LIF), as well as electron density and electron temperature by a multipole resonance probe (MRP). A simple collisional-radiative model for UV emission from NO(A) is developed, which takes the measured densities of ground state nitric oxide and electrons, as well as the electron temperature and neutral gas temperature, as input parameters. The results reveal the links between the absolute densities of ground state nitric oxide, the excitation of this species driven by electron impact and collisions with nitrogen metastables, quenching of the nitrogen metastables, and the resulting UV photon emission rate. The density of ground state nitric oxide is shown to increase with power, while the discharge remains in E-mode, and to decrease significantly with the transition into H-mode, when sufficient rf power is deposited in the discharge. Despite the lower densities of ground state nitric oxide in H-mode, the UV photon emission intensity increases continuously with higher rf powers and over the E-H transition. This effect is shown to be caused by increased excitation of NO(A) by nitrogen metastables in H-mode, which is sufficient to overcompensate the decrease in ground state nitric oxide density.
\end{abstract}



\vspace{1pc}
\end{adjustwidth}



\ioptwocol

\section{Introduction}
\label{chap:introduction}
Inductively coupled low-pressure plasmas (ICP) are a type of plasma device that are essential to, or investigated for, many surface treatment applications, such as nanoscale processing in semiconductor manufacturing, sterilization and disinfection in biomedicine, and surface modification of sensitive substrates \cite{hopwood1992, keller1997, lee2018, halfmann2007, gans2005}. ICPs are typically operated at radio frequency (RF) and differ from the also commonly used capacitively coupled plasmas (CCP) in two key aspects. Under comparable operating conditions and power input they offer higher electron densities and lower ion energies, yielding the opportunity to produce higher densities of reactive species or photon emission, while avoiding high-energy ion bombardment of the substrate. Depending on the application, these reactive species and photons, particularly those in the ultraviolet (UV) and vacuum ultraviolet (VUV) ranges, enable a variety of effective surface treatments. In addition, ICPs offer two modes of operation, the capacitive E-mode, which occurs at lower power coupling, and the inductive H-mode, at higher power coupling, which gives the discharge its name. The ability to switch between these two modes at will, e.g. by adjusting the supplied rf power, offers opportunities to directly affect the balance of neutral, excited, and charged species, as well as photon emission and energy, within the discharge. \cite{lieberman, chabert}. Further, uncontrolled transitions between the two modes due to instabilities \cite{corr2003,chabert2001}, or time dependent mode transitions during pulsed operation can also lead to challenges in process control \cite{qu_power_2020, piskin2023}, and as a result it is important to understand the basic properties of plasmas operating in both E- and H-mode. \\

The properties of ICP systems across the E-H mode transition have been widely investigated by experiment and modelling in the literature, and are of significance for a number of applications. Experimental studies have been carried out in a variety of gas mixtures and have focused on a range of plasma and system properties, such as the electron densities \cite{ostrikov_power_2000, malyshev_diagnostics_2000_2, corr_comparison_2008, corr_discharge_2012, fei_changes_2013, liu_mode_2013, ahr_influence_2015, wegner_electron_2015, wegner_e-h_2017}, electron temperatures \cite{ostrikov_power_2000, malyshev_diagnostics_2000_2,  fuller_characterization_2000, corr_comparison_2008, corr_discharge_2012, wegner_e-h_2017}, electron heating phenomena \cite{zaka-ul-islam_energetic_2011, fei_changes_2013, ahr_influence_2015, wegner_electron_2015}, negative ion formation \cite{corr_comparison_2008, wegner_e-h_2017_2}, gas temperatures \cite{wegner_e-h_2017, meehan_gas_2020} and reactive species formation \cite{fuller_characterization_2000, malyshev_diagnostics_2000, corr_comparison_2008, corr_discharge_2012, wegner_e-h_2017, zeng_spatially_2019}. Recent reviews provide comprehensive overviews of current knowledge relating to the physics and applications of ICPs, including their properties across the E-H mode transition \cite{lee_review_2018, mitsui_review_2021}. Here we add to this body of work through an extensive experimental characterisation of a double-inductively coupled plasma (DICP) system, complemented by plasma-chemical modelling, to advance understanding of the relationship between different plasma parameters in ICPs formed in nitrogen-oxygen mixtures. \\

The DICP system used in this work has been studied extensively in gas compositions of argon mixed with molecular gases, especially with respect to fundamental plasma parameters such as gas temperature, and photon emission, as functions of gas pressure and rf power \cite{fiebrandt2017, fiebrandt2018, fiebrandt2020, diss_fiebrandt2018, engelbrecht2024}. The diagnostics used in these past studies include, amongst others, Langmuir probe (LP) and multipole resonance probe (MRP), absolutely calibrated optical emission spectroscopy (OES), and absorption spectroscopy. In this work, an extensive parameter scan of pressure, power, and gas composition is performed on the system, with gas mixtures consisting of nitrogen and oxygen. This gas mixture enables the direct production of nitric oxide, which has been shown to be a molecule of high interest in sterilization and semiconductor processing due its inherent reactive nature, and as the primary source of UV photon emission in such discharges \cite{laroussi2005, fiebrandt2018_patent, kogelheide2020, kutasi2008}. Specifically, UV radiation from NO can lead to fast inactivation of microorganisms for sterilization \cite{fiebrandt2018}, and NO molecules can play a role in the etching mechanism of certain semiconductor processing steps \cite{huang_downstream_2018, kastenmeier1998}. In addition, UV radiation in the range corresponding to that emitted by excited NO molecules can lead to defect formation in material stacks relevant for semiconductor manufacturing \cite{miyoshi_reduction_2020, fukumizu_atomic_2019}. Laser-induced fluorescence (LIF) is performed on the DICP system, in order to measure absolute densities of ground state NO molecules. These densities of ground state NO are also the basis for the complementary plasma-chemical modelling, using a collisional radiative model. Both LIF and MRP use the gas temperatures determined by OES as input parameters, which increases their reliability across the entire parameter range and over the E-H mode transition. \\

The UV radiation produced by the discharge is, for the most part, produced by excited states of nitric oxide \cite{kutasi2008,fiebrandt2018}. This means that UV emission will depend on the absolute density of nitric oxide in the ground state, and the production of nitric oxide excited states based on the electron energy distribution function (EEDF), over the given parameter range. The E-H mode transition, which is strongly dependent on the gas pressure, gas composition, and rf power, causes a significant change in gas temperature and electron density. This affects the density of nitric oxide and the corresponding UV emission in a non-linear way, as shown by the results in this work. This, in turn may provide opportunities to tailor comparable ICP systems to control for effects related to NO molecules and UV radiation, depending on the intended application. \\

\section{Experiment and diagnostics}
\label{chap:experiment}

\subsection{Double-inductively coupled plasma system}
\label{chap:dicp}
The experimental setup used for all experiments shown in this work is a double-inductively coupled plasma (DICP) system. The system consists of a cylindrical stainless steel chamber with an inner diameter of 40\,cm and a height of 20\,cm. The bottom and lid of the system are made from borosilicate glass, sealed against the steel cylinder by fluorocarbon (FKM) o-ring seals. The lid is attached to the chamber via a hinge, enabling ease of access to the inside of the chamber, e.g. for placing biological samples on an optional sample holder. Two in-series vacuum pumps, a roots pump (EH 500, Edwards Vacuum Ltd, Burgess Hill, England) and a rotary vane pump (DUO 060 A, Pfeiffer Vacuum Technology AG, Aßlar, Germany), are used to generate a base vacuum of 0.1\,Pa in the chamber. An optional turbo pump is also connected to the setup but not used in this work. The typical leak rate of the system is in the order of 0.03\,sccm. The pressure of the system is monitored by three pressure gauges for different ranges. Up to 100\,Pa and up to 10\textsuperscript{5}\,Pa: MKS 627B (MKS Instruments Inc., Andover, USA); Full-range: PKR 250 (Pfeiffer Vacuum Technology AG, Aßlar, Germany). A downstream butterfly valve (MKS 653B, MKS Instruments Inc., Andover, USA) is used for controlling the pressure in the range between 1\,Pa and 80\,Pa. \\

Optical access to the discharge is facilitated by several KF40 flanges and optical windows made from quartz or borosilicate glass, depending on the requirements of the application. The KF40 flanges are also used for the access of probe diagnostics, such as the MRP, in this work. Process gas is supplied by several mass flow controllers (multiple models) and includes argon, nitrogen, oxygen, hydrogen, and other gases, such as calibration gas mixtures, as required (typical purity grade N5.0, Alphagaz 1, Air Liquide S.A., Paris, France). \\

\begin{figure}[t!]
    \centering
    \includegraphics[width=0.495\textwidth]{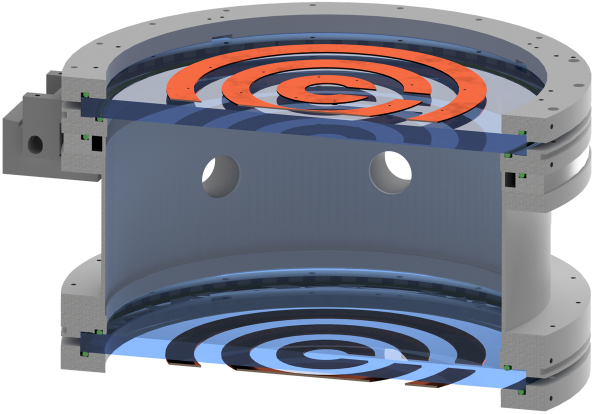}
	\caption{Digital sketch of the double-inductively coupled plasma (DICP) system used for the experimental results presented in this work. The chamber consists of a cylindrical stainless steel vacuum chamber, with lid and bottom made from borosilicate glass. The rf power at 13.56\,MHz is supplied by a commercial power generator and a custom matchbox (both not shown here), which are connected in phase to the two planar copper coils, resting on the glass discs outside of vacuum. This figure is redrawn and adapted based on a similar sketch published in \cite{fiebrandt2018} © IOP Publishing Ltd. All rights reserved.}
    \label{fig:dicp}
\end{figure}

The rf power at 13.56\,MHz is provided by two symmetrical, copper planar coil antennas, located on the outside of the glass lid and bottom of the chamber, respectively. A simplified, true to scale sketch of the system is shown in figure~\ref{fig:dicp}. The two antennas are connected in phase to a custom rf matchbox and rf power generator (Dressler CESAR 1350D, Advanced Energy Industries Inc., Denver, USA), used for power delivery up to 800\,W, but generally capable of delivering 5000\,W. The matchbox provides capability for manual and automatic matching, where the former if used for ignition of the plasma in E-mode, and the latter for automatic optimization and control of power delivery during the subsequent process. Reflected powers are generally below 1\,\%, once the system is in steady state. Typically, the matching system achieves that steady state within a few seconds and is capable of smoothly transitioning from E-mode to H-mode. \\

The entire experimental system, with exception of the external diagnostic methods, is controlled by a single LabVIEW control software (National Instruments Corp., Austin, USA), ensuring reproducibility of the performed experiments. Evaluation and post-processing of the measurements is performed externally in the respective software of the diagnostic systems (MRP), in Julia language (OES, LIF), and visualized using the open-source ``PlotlyJS.jl" library \cite{plotly}. \\

\subsection{Laser-induced fluorescence}
\label{chap:lif}
Laser-induced fluorescence (LIF) is an optical diagnostic used to measure absolute densities of specific molecules in gaseous systems. LIF makes use of the excitation of molecules in the state to be measured to a higher energy state, and the subsequent fluorescent emission of light from the higher state. The exciting laser pulse of light usually has a width in the order of a few nanoseconds, so that a defined exponential decay of the fluorescence signal, over a few dozens of nanoseconds, may be detected and evaluated. Typically, a calibration gas of a known concentration is required to acquire absolute densities of the measured species. \\

For the detection of nitric oxide (NO) in this study, a dye laser (Narrowscan, Radiant Dyes Laseraccessoires GmbH, Wermelskirchen, Germany) is used to excite the NO molecules within the plasma. The dye laser is pumped by a frequency tripled Nd:YAG laser (Spitlight Compact DPSS, InnoLas Laser GmbH, Krailling, Germany) and makes use of the dye Coumarin-47, so that the laser wavelength may be tuned from 442\,nm to 479\,nm. To excite the relevant energy levels of nitric oxide the laser frequency is doubled by a barium borate (BBO) crystal. The final LIF excitation wavelength, used for all such measurements in this work, is 226.3\,$\pm$\,0.2\,nm, corresponding to the excitation of the NO(A, $v=0$) from the ground state NO(X, $v=0$). The considered emission, caused by fluorescent transition from the NO(A, $v=0$) to the state NO(X, $v=2$), is in the wavelength range of around 245\,nm to 248\,nm. The laser pulse energies are measured to be approximately 130\,µJ over 5\,ns, at pulse repetition rates of 100\,Hz, and with a beam diameter of around 200\,µm. \\

To ensure reliable and reproducible measurements, the laser energy was tracked by a photodiode (SM05PD7A, Thorlabs Inc., Newton, USA) during all experiments, and the measured signal intensity corrected accordingly. This was achieved by a beam splitter placed at the outlet of the laser system, directing a part of the laser beam onto the photodiode. Furthermore, saturation effects of the LIF signal were not observed, as the signal remained linear over broad scans of the laser energy, exceeding the used operational range. The laser beam was then focused into the center of the reactor chamber by a quartz lens of 300\,mm focal length (LB4545, Thorlabs Inc., Newton, USA). The LIF emission occurring in the focal point of the laser beam was refocused onto the detector by another lens of 200\,mm focal length (LB4842, Thorlabs Inc., Newton, USA), placed diagonally to the incident beam and in a 2f setup, to reduce the length of the optical path. To isolate the fluorescence signal from the total light emission, the refocused light went through an optical filter of 248\,nm center wavelength (\#67-806, Edmund Optics Inc., Barrington, USA), and was then transmitted onto a photo multiplier tube (R928, Hamamatsu Photonics K.K., Hamamatsu, Japan) for detection. In combination with an oscilloscope of 2.5\,GHz bandwidth (DPO 7254C, Tektronix Inc. Beaverton, USA), the use of a fast photo multiplier tube ensured a sufficient temporal resolution to accurately integrate and evaluate the fluorescence signal. \\

The system was calibrated using a test gas of \mbox{5\,\% $\pm$ 2\,\% rel.} of nitric oxide in helium of N5.0 purity (CRYSTAL-Mixture, Air Liquide S.A., Paris, France) and a linear interpolation curve at a constant pressure and by dilution with argon. Initial attempts used additional helium to further dilute the calibration gas mixture, however, this process led to issues that were ascribed to unmixing of the gas mixture. Using argon as the diluting gas alleviated this issue and resulted in temporally stable and linear calibration signals. Overall, the calibration curve covered an NO density range from $1.53 \times 10^{17}$\,m\textsuperscript{-3} to $1.23 \times 10^{20}$\,m\textsuperscript{-3}, with a total of 10 calibration values. The resulting linear interpolation curve was chosen with a y-intercept at zero and resulted in a mean error of 1.04\,\%. With the calibration being acquired at room temperature, higher gas temperatures during measurement need to be taken into account to acquire reliable densities by LIF. Here, the vibrational and rotational temperatures of NO(A) are assumed to be identical to the gas temperature measured by OES (see following section). The change in the rotational and vibrational distributions affects the fluorescence intensity of the considered 226.3\,nm line, by altering the transition probabilities. We account for these effects by simulating the relative difference in intensity of the lines of interest in the LIFBASE software, for the measured differences in gas temperature \cite{lifbase}. These simulations in LIFBASE also show that the temperature induced changes of the rotational state distribution are roughly ten times as important as those of the vibrational state distribution, for gas temperatures of up to 1000\,K. Broadening effects of the line are taken to be negligible. Reproducibility of the measurement procedure has been regularly checked by repetition of the calibration procedure and no significant deviations were observed. \\

\subsection{Optical emission spectroscopy}
\label{chap:oes}
Optical emission spectroscopy (OES) is a non-invasive diagnostic method highly suitable for plasmas, due to their inherent emission of light. The light emitted by a plasma is characteristic for the collisional-radiative processes occurring within the discharge, and the underlying plasma chemistry. Multiple parameters of a plasma discharge may be determined from OES, by analysis of individual emission line ratios, their broadening, or their absolute intensities. Such parameters, under application of suitable theoretical models, include, amongst others, the gas and electron temperatures, as well as densities of electrons, excited species, ions, and neutral species. In this work OES is used specifically for the determination of the gas temperature of the discharge, based on the comparison of rotational structures of molecular nitrogen emission bands with corresponding simulated spectra, as well as absolute UV photon emission rates, which are attributed to excited states of nitric oxide. \\

The corresponding emission spectra in this work are acquired by use of a relatively and absolutely calibrated echelle spectrometer (ESA4000, LLA Instruments GmbH \& Co. KG, Berlin, Germany). The calibration procedure is performed in accordance with the method described by Bibinov et al., using an identical echelle spectrometer and two absolutely calibrated light sources: a deuterium lamp for the wavelength region of 200\,nm to 350\,nm and a tungsten-ribbon lamp for the region of 350\,nm to 900\,nm \cite{bibinov2007}. Here, emission spectra are acquired by use of an optical fibre placed directly in front of one of the chambers quartz windows, and the acceptance angle is restricted by an aperture with an acceptance angle of $2\,\alpha=5.8$\,\textdegree. The solid angle over which the photons were accumulated during measurement was directed horizontally through the center of the discharge, and verified by use of an LED attached to the fibre in the reverse direction. The gate width of the spectrometer was set to a constant 5\,ms with spectra being accumulated to the point of achieving a consistent intensity below the CCD chips threshold for over-exposure. Background spectra were acquired for each measurement series and, corrected to the respective exposure time for each individual measurement, and subtracted from each measurement. \\

For the evaluation of gas temperatures the N\textsubscript{2}(C\textsuperscript{3}\textPi\textsubscript{u}) to N\textsubscript{2}(B\textsuperscript{3}\textPi\textsubscript{g}) emission band at 337\,nm is analyzed by comparison to simulated spectra and was demonstrated by Bibinov et al. \cite{bibinov2011}. Each measured spectrum is compared manually against a set of simulated spectra of the band, over a temperature variation in increments of 20\,K. In particular, the rotational structure of the band from around 334\,nm to 336\,nm is compared against the simulated spectra, with the gas temperature being determined by the best fit. The gas temperatures determined by this method were then used as one of the input parameters to the other diagnostics applied in this work. It should be emphasised that inferring the gas temperature by this method assumes that the translational temperature of the gas is in equilibrium with the rotational temperature of the respective transition \cite{bibinov2011}. \\

With respect to UV photon emission, the absolutely calibrated spectra acquired by the aforementioned method were integrated numerically over their respective wavelength range using a trapezoidal method. The UV bands used for this are defined as follows: 380\,-\-315\,nm for \mbox{UV-A}, 315\,-\-280\,nm for \mbox{UV-B}, and 280\,-\-200\,nm for \mbox{UV-C}. Technically, \mbox{UV-C} is defined from 280\,-\-100\,nm, however, the used echelle spectrometer and quartz windows are limited at 200\,nm for the lower wavelength range. All UV photon emission rates given in this work are the sum of the total 380\,-\-200\,nm range. Emission in the lower \mbox{UV-C} range is dominated by the \textgamma-band with the NO(A\textsuperscript{2}\textSigma\textsuperscript{+}) to NO(X\textsuperscript{2}\textPi) transition, while emission in the higher \mbox{UV-C} range is caused by a combination of the former band and the \textbeta-band with the NO(B\textsuperscript{2}\textPi) to NO(X\textsuperscript{2}\textPi) transition. In the \mbox{UV-B} and \mbox{UV-A} range only the \textbeta-band dominates the UV emission. Absolute values for the individual ranges have been determined but serve no additional benefit for the analysis of the results here, and are not shown. \\

\subsection{Multipole resonance probe}
\label{chap:mrp}
The multipole resonance probe (MRP), belongs to the class of probe diagnostics that are usually placed within the plasma discharge, by means of a vacuum feed-through. Such probes are often equipped with vacuum gaiters and motors, enabling linear movement of the probe, in order to scan the measured parameters across different sections of the plasma discharge. \\

The MRP, which basically acts as a high-frequency antenna, relies on an extensive theoretical plasma model and electrical network theory, to derive fundamental plasma parameters from a discharge. An overview of the underyling theoretical concepts is given in \cite{gong2022,oberrath2021,oberrath2014,lapke2008}. A network analyzer is used to apply a high-frequency sweep to the MRP, which transmits that frequency sweep into the discharge volume. The network analyzer can then evaluate the coupling between MRP and discharge for the input port voltage reflection coefficient S\textsubscript{11}. This reflection coefficient contains signals correlated to the absorption of parts of the frequency sweep by the charges within the plasma, primarily the electrons. By use of the aforementioned plasma theory, equations for the electron temperature \textit{n}\textsubscript{e} and electron temperature \textit{T}\textsubscript{e} can be derived from the measured S\textsubscript{11} coefficient and a known neutral gas temperature \textit{T}\textsubscript{g}. \\

The spherical multipole resonance probe used in this work (MRP PM, House of Plasma GmbH, Bochum, Germany) has a diameter of 5\,mm and was placed, within a borosilicate glass tube of 8\,mm diameter, at a fixed position in the center of the discharge. This position coincides with the position where the signal of the LIF measurements was acquired, to ensure the highest possible comparability. The system was used in conjunction with additional hardware and software provided by the manufacturer, which includes extensive capabilities for evaluation of the measurements, yielding values for \textit{n}\textsubscript{e} and \textit{T}\textsubscript{e}, without additional post-processing. The required gas temperatures were given, as illustrated in section~\ref{chap:oes}, by optical emission spectroscopy. The determined plasma parameters then also served as input parameters for the chemical kinetics model, as described in chapter~\ref{chap:simulation}. \\

\subsection{Error estimation}
\label{chap:errors}
\subsubsection{Presentation of estimated errors}
The estimated errors for experimental data presented in this work are determined as the relative errors for each respective diagnostic, based on the underlying uncertainties of the used values from literature and the diagnostic devices. Statistical errors are omitted, as they are likely to be much smaller than the systematic errors, due to the good reproducibility of the experimental setup and the diagnostic methods, as shown in previous studies. These systematic errors are discussed in this section and given in the captions of the corresponding figures, rather than as error bars within the figures, for the sake of readability. \\

\subsubsection{Uncertainty of the absolutely calibrated emission spectra}
For the absolutely calibrated optical emission spectroscopy (OES), which is used to determine the UV photon emission rates, the error is determined not just by the measurement procedure itself, but also by the quality of the absolute calibration. The calibration of the used Echelle spectrometer is carried out in regular intervals of at least 6 months, and re-checked when the device is transported or exposed to significant temperature changes. Ultimately, the error of the UV spectra presented in this work is assumed to be mainly defined by the standard error of the absolute calibration procedure, which was determined as 12\,\% by Bibinov et al. \cite{bibinov2007}. In comparison to that, inaccuracies of the geometrical parameters, going into the equation for absolute calibration of the spectra, are assumed to be negligible. \\

\subsubsection{Uncertainty of the gas temperatures}
With respect to the gas temperatures, as also determined by OES, the error is defined by the accuracy with which the rotational bands of the N\textsubscript{C-B} band at 337\,nm can be fitted to the simulated spectra, for each respective temperature in steps of 20\,K (for additional details see chapter \ref{chap:oes}). Especially at lower temperatures this fitting procedure is difficult to perform numerically and, thus, is performed manually instead. A typical confidence interval using this manual procedure is in the order of 40\,K for the entire temperature range. In the worst possible scenario, where a plasma is close to 300\,K, these 40\,K uncertainty amount to a relative error of roughly 13.5\,\%. This confidence interval is then attributed to all measured temperatures given in this work, even at higher temperatures, where the overall error might be smaller. \\

\subsubsection{Uncertainty of the laser-induced fluorescence}
The systematic error of the laser-induced fluorescence, as used in this work, depends on three major factors. The uncertainty of the nitric oxide concentration in the calibration gas, the accuracy of the detector system, and the uncertainty of the gas temperature, which has a significant impact on the intensity of the fluorescence signal and is taken into account in the evaluation procedure. Possible changes in the intensity of the exciting laser beam are tracked by a photodiode and automatically corrected for in the evaluation procedure. The confidence interval for the concentration of NO in the calibration gas is specified by the manufacturer as 2\,\%. The detector system relies on the accuracy of the oscilloscope, which is 1.5\,\% in the used range. The most significant factor is the uncertainty of the gas temperature of up to 13.5\,\%, as described in the previous section. In the worst possible case this deviation of the gas temperatures translates into an error of 20.6\,\% for the emission intensity of the excited NO states after laser excitation, as determined by simulations using the LIFBASE software \cite{lifbase}. Overall, the root of the sum of the squares for these errors results in a total interval of confidence for the NO densities acquired by LIF of 21\,\%. \\

\subsubsection{Uncertainty of the multipole resonance probe}
In the case of the multipole resonance probe, it is more challenging to define an overall confidence interval, due to the underlying theory of the diagnostic. The diagnostic itself is, due to the good definition of its electronic components, highly reproducible and the statistical error rather negligible. Instead the major uncertainty is defined by the underlying physical model that is used to calculate the electron temperature and density from the plasma admittance, and its inherent assumptions. For cases that differ more strongly from the conditions assumed for that model, the deviations are expected to get increasingly higher, but can not easily be quantified. In a study performed by Fiebrandt et al. on the same system as used in this work, the MRP was compared against the well established Langmuir probe (LP) method for different operating conditions of argon in variable molecular gas mixtures \cite{fiebrandt2017}. Fiebrandt et al. report a typical deviation in the order of 10\,\% between electron densities measured by MRP and those measured by LP \cite{fiebrandt2017}. However, towards lower fractions of argon and higher fractions of molecular gases they also report increasing deviations between MRP and LP. This is particularly the case for oxygen admixture, where, in the case of pure oxygen, the maximum discrepancy between the two probes is in the order of 80\,\%. Overall, Fiebrandt et al. conclude that the MRP is expected to work well in molecular gasses, but requires additional validation \cite{fiebrandt2017}. As a consequence, it is challenging to quantify a specific interval of confidence for the MRP, especially with the mixture of nitrogen and oxygen used in this work. Nevertheless, the results are expected to be in the correct order of magnitude, while also reflecting qualitative trends accurately. \\

\section{Simple collisional-radiative model}
\label{chap:simulation}

\renewcommand{\arraystretch}{1.2}
\begin{table*}[htbp]
    \caption{System of collisions and radiative processes considered in the zero-dimensional model for the determination of UV emission at 236\,nm from the radiative decay of the first excited state of nitric oxide NO(A, $v=0$). Energy units: Electron temperature \textit{T}\textsubscript{e} [eV]; Gas temperature \textit{T}\textsubscript{g} [K]. Energy dependent coefficients: Analytical function f(\textit{T}\textsubscript{e,g}); Interpolated function f\textsubscript{{\normalfont i}}(\textit{T}\textsubscript{e,g}). Rate coeff. units: Single body reaction [s\textsuperscript{-1}]; Two-body reaction [m\textsuperscript{3}s\textsuperscript{-1}]. Where no \textit{T}\textsubscript{g} dependence is given for a reaction, values refer to those determined at 300 K, and are likely to be less accurate at the higher gas temperatures found under a range of conditions in this work. \\}
    \begin{tabular}{@{}lllc@{}}
        \toprule
        No. & Reaction & Rate constant & Ref./Notes \\
        \midrule
        R1 & \ce{N2(X) + e -> N2(A) + e} & $f_{\mathrm{i}}(T_{\mathrm{e}})$ & \cite{itikawa2006} \\
        R2 & \ce{N2(X) + e -> N2(B) + e} & $f_{\mathrm{i}}(T_{\mathrm{e}})$ & \cite{itikawa2006} \\
        R3 & \ce{N2(X) + e -> N2(C) + e} & $f_{\mathrm{i}}(T_{\mathrm{e}})$ & \cite{itikawa2006} \\[0.4em]
        
        R4 & \ce{N2(A) + e -> N2(B) + e} & $f_{\mathrm{i}}(T_{\mathrm{e}})$ & \cite{bacri1981} \\
        R5 & \ce{N2(A) + e -> N2(C) + e} & $f_{\mathrm{i}}(T_{\mathrm{e}})$ & \cite{bacri1981} \\
        R6 & \ce{N2(B) + e -> N2(C) + e} & $f_{\mathrm{i}}(T_{\mathrm{e}})$ & \cite{bacri1981} \\[0.4em]
        
        R7 & \ce{N2(C) -> N2(B) + h\nu} & $2.74 \times 10^{7}$ & \cite{N2_C-B,guerra1997} \\
        R8 & \ce{N2(B) -> N2(A) + h\nu} & $2.00 \times 10^{5}$ & \cite{N2_B-A, N2_B-A_2, guerra1997} \\[0.4em]
        
        R9 & \ce{N2(A) + wall -> N2(X)} & $f(T_{\mathrm{g}})$ & \cite{gudmundsson2009}$^{a}$ \\
        R10 & \ce{N2(A) + O2 -> N2(X) + O2} & $8.75 \times 10^{-19}(T_{\mathrm{g}}/300)^{0.55}$ & \cite{guerra1997,N2A+O2_1,N2A+O2_2,N2A+O2_3} \\[0.4em]
        
        R11 & \ce{N2(A) + N2(A) -> N2(B) + N2(X)} & $7.70 \times 10^{-17}$ & \cite{N2A+N2A_N2B} \\
        R12 & \ce{N2(A) + N2(A) -> N2(C) + N2(X)} & $1.50 \times 10^{-16}$ & \cite{N2A+N2A_N2C} \\[0.4em]
        
        R13 & \ce{N2(B) + N2(X) -> N2(A) + N2(X)} & $2.85 \times 10^{-17}$ & \cite{guerra1997, guerra1997_2, N2B+N2_N2A+N2} \\
        R14 & \ce{N2(B) + N2(X) -> N2(X) + N2(X)} & $0.15 \times 10^{-17}$ & \cite{guerra1997, guerra1997_2} \\[0.4em]
        
        R15 & \ce{N2(A) + O -> NO(X) + N(^2D)} & $7.00 \times 10^{-18}$ & \cite{guerra1995,N2A+O_N2X+O,piper1981} \\
        R16 & \ce{N2(A) + O -> N2(X) + O(^1S)} & $2.10 \times 10^{-17}$ & \cite{N2A+O_N2X+O} \\
        R17 & \ce{N2(A) + O2 -> N2(X) + O + O} & $1.63 \times 10^{-18}(T_{\mathrm{g}}/300)^{0.55}$ & \cite{guerra1997,N2A+O2_1,N2A+O2_2,N2A+O2_3} \\
        R18 & \ce{N2(B) + O2 -> N2(X) + O + O} & $3.00 \times 10^{-16}$ & \cite{guerra1997,kossyi1992} \\
        R19 & \ce{N2(C) + O2 -> N2(X) + O + O} & $3.00 \times 10^{-16}$ & \cite{guerra1997,kossyi1992} \\[0.4em]
        
        R20 & \ce{N2(B) + NO(X) -> N2(A) + NO(X)} & $2.40 \times 10^{-16}$ & \cite{N2B+NO_N2A+NO} \\[0.4em]
        
        R21 & \ce{NO(X) + e -> NO(A, $v=0$) + e} & $f_{\mathrm{i}}(T_{\mathrm{e}})/0.277$ & \cite{NO_X-A,NO_lifetime}$^{b}$ \\
        R22 & \ce{NO(X) + N2(A) -> NO(A, $v=0$) + N2(X)} & $6.60 \times 10^{-17}$ & \cite{NO_X-A_byN2a} \\[0.4em]
        
        R23 & \ce{NO(A, $v=0$) -> NO(X) + h\nu} & $5.19 \times 10^{6}$ & \cite{NO_lifetime}$^{c}$ \\
        R24 & \ce{NO(A, $v=0$) + N2(X) -> NO(X) + N2(X)} & $1.074\times 10^{-19}(T_{\mathrm{g}}/300)^{2.79} e^{-664.710/T_{\mathrm{g}}}$ & \cite{qNO_byN2} \\
        R25 & \ce{NO(A, $v=0$) + O2 -> NO(X) + O2} & $2.549\times 10^{-16}(T_{\mathrm{g}}/300)^{0.62} e^{21.763/T_{\mathrm{g}}}$ & \cite{qNO_byO2} \\
        R26 & \ce{NO(A, $v=0$) + NO(X) -> NO(X) + NO(X)} & $2.00 \times 10^{-16}$ & \cite{pintassilgo2005,zacharias1976} \\[0.4em]
        
        R27 & \ce{O2 + e -> O + O + e} & $f_{\mathrm{i}}(T_{\mathrm{e}})$ & \cite{alves2014, lxcat_lisbon}$^{d}$ \\ 
        R28 & \ce{O + wall -> 1/2O2} & $f(T_{\mathrm{g}})$ & \cite{gudmundsson2007}$^{e}$ \\[0.4em]

        
        R29 & \ce{N2(X) + e -> N + N + e} & $f_{\mathrm{i}}(T_{\mathrm{e}})$ & \cite{bacri1981} \\
        R30 & \ce{N2(A) + e -> N + N + e} & $f_{\mathrm{i}}(T_{\mathrm{e}})$ & \cite{bacri1981} \\
        R31 & \ce{N2(B) + e -> N + N + e} & $f_{\mathrm{i}}(T_{\mathrm{e}})$ & \cite{bacri1981} \\
        R32 & \ce{N2(C) + e -> N + N + e} & $f_{\mathrm{i}}(T_{\mathrm{e}})$ & \cite{bacri1981} \\
        
        \bottomrule
    \end{tabular} \\ \\
    \footnotesize
   $^{a}$ Calculated using the expression for surface losses given in the cited reference with a surface loss coefficient of $\gamma=1$. \\
   $^{b}$ The emission cross section for electron impact from NO(X), leading to the NO(A\textsuperscript{2}\textSigma\textsuperscript{+}, $v=0$\,$\rightarrow$\,X\textsuperscript{2}\textPi, $v=1$) transition from \cite{NO_X-A} is used. Since the aim is to simulate the density of the NO(A, $v=0$) state, the emission cross section is divided by the branching ratio for the specific transition \cite{NO_lifetime} in order to estimate the total excitation into the NO(A, $v=0$) state from the emission cross section. \\ 
   $^{c}$ Here, the total Einstein coefficient for decay of the NO(A, $v=0$) level is used to simulate the NO(A, $v=0$) density. The fraction of emission at 236 nm is calculated by multiplying the rate for this process by the corresponding branching ratio.  \\
   $^{d}$ Electron impact cross sections with energy thresholds of 6 eV and 8.4 eV, leading to the formation of O($^3$P)+O($^3$P) and O($^3$P)+O($^1$D), respectively, are included. O($^3$P) and O($^1$D) are treated as equivalent and included in the density of O. \\
    $^{e}$ Calculated using the expression for surface losses given in the cited reference with a surface loss coefficient of $\gamma=0.154$ \cite{gudmundsson2007}. \\
    \label{tab:reactions}
\end{table*}

Simulations of reaction kinetics and radiation emission in plasma sources can allow for deeper insights into the process by successive simplification of the model and identification of the major driving mechanisms for the observed properties. In this study such a simplified model of rate equations in zero spatial dimensions is solved on the basis of electron temperatures and densities measured by MRP, which would otherwise have to be simulated in an additional plasma model, as well as the gas temperature measured by OES and the ground state NO density measured by LIF. The major aim of the model is to better understand the pathways towards UV emission from the excited state NO(A) across the E-H mode transition, and how these relate to the other plasma parameters. The used reaction system is given in table~\ref{tab:reactions}. To avoid simulating all emission lines from excited NO, the simplified collisional-radiative model is focused on the formation of the NO(A, $v=0$) state, followed by emission around 236 nm i.e. the NO(A\textsuperscript{2}\textSigma\textsuperscript{+}, $v=0$\,$\rightarrow$\,X\textsuperscript{2}\textPi, $v=1$) transition of the NO \textgamma-system. The scheme presented by Offerhaus et al. \cite{offerhaus2019} for atmospheric pressure discharges is used as starting point for this work. More sophisticated models of Guerra et al. and Gudmundsson et al. are then used to identify additional important reactions and expand the scheme, with the aim of achieving better quantitative agreement between simulation and measurement \cite{guerra2001, gudmundsson2007}. Overall, formation of NO(A) is assumed to occur via electron impact excitation from the NO ground state as well as via collisions with excited N\textsubscript{2} molecules, specifically N\textsubscript{2}(A). The majority of the reaction scheme is aimed towards including important formation and consumption processes of N\textsubscript{2}(A), and therefore includes a number of other species, such as N\textsubscript{2}(B), N\textsubscript{2}(C), O, that have been found to be important for this in the literature. \\

For reactions involving electron impact, a Maxwellian electron energy distribution function at the respective electron temperature \textit{T}\textsubscript{e}, together with the cross-section for the process (see column. 'Ref.'), are used to calculate rate constants. The rate for diffusion losses of atomic oxygen (R21) and N\textsubscript{2}(A) (R9) to surfaces are given by analytical functions of the gas temperature \textit{T}\textsubscript{g}, derived by Gudmundsson et al. \cite{gudmundsson2007, gudmundsson2009}. The model requires the measured densities of electrons and ground state NO, as well as the densities of N\textsubscript{2}, O\textsubscript{2}, given by the pre-defined gas mixture, as input parameters. The model is then solved for the temporal evolutions of the remaining species, which are N\textsubscript{2}(A), N\textsubscript{2}(B), N\textsubscript{2}(C), O, and NO(A, $v=0$). The resulting reaction set is solved using the ``Catalyst.jl" package in Julia language \cite{CatalystPLOSCompBio2023}. For all cases, convergence is reached after less than one second of simulated time, as is characteristic for the fast dynamics of low-pressure plasmas. To compare measurement and simulation, the NO emission band around 236\,nm is used. The emission intensity at this wavelength is calculated from the model by multiplying the total decay rate of the NO(A, $v=0$) state (R21) with the branching ratio of 0.277 for emission at 236\,nm \cite{NO_lifetime}. \\

\section{Results and discussion}
\label{chap:results}

\subsection{Experimental results}
\label{chap:experimental_results}
Parameter variations in inductively coupled discharges are especially of interest with respect to the E-H mode transition and its implications for different properties of the discharge. In this study, the rf power applied to the system is the main lever to initiate this transition, while a broad selection of gas pressures illustrates the sensitivity of the measured parameters to one another. \\

\begin{figure}[t!]
    \centering
    \includegraphics[width=0.495\textwidth]{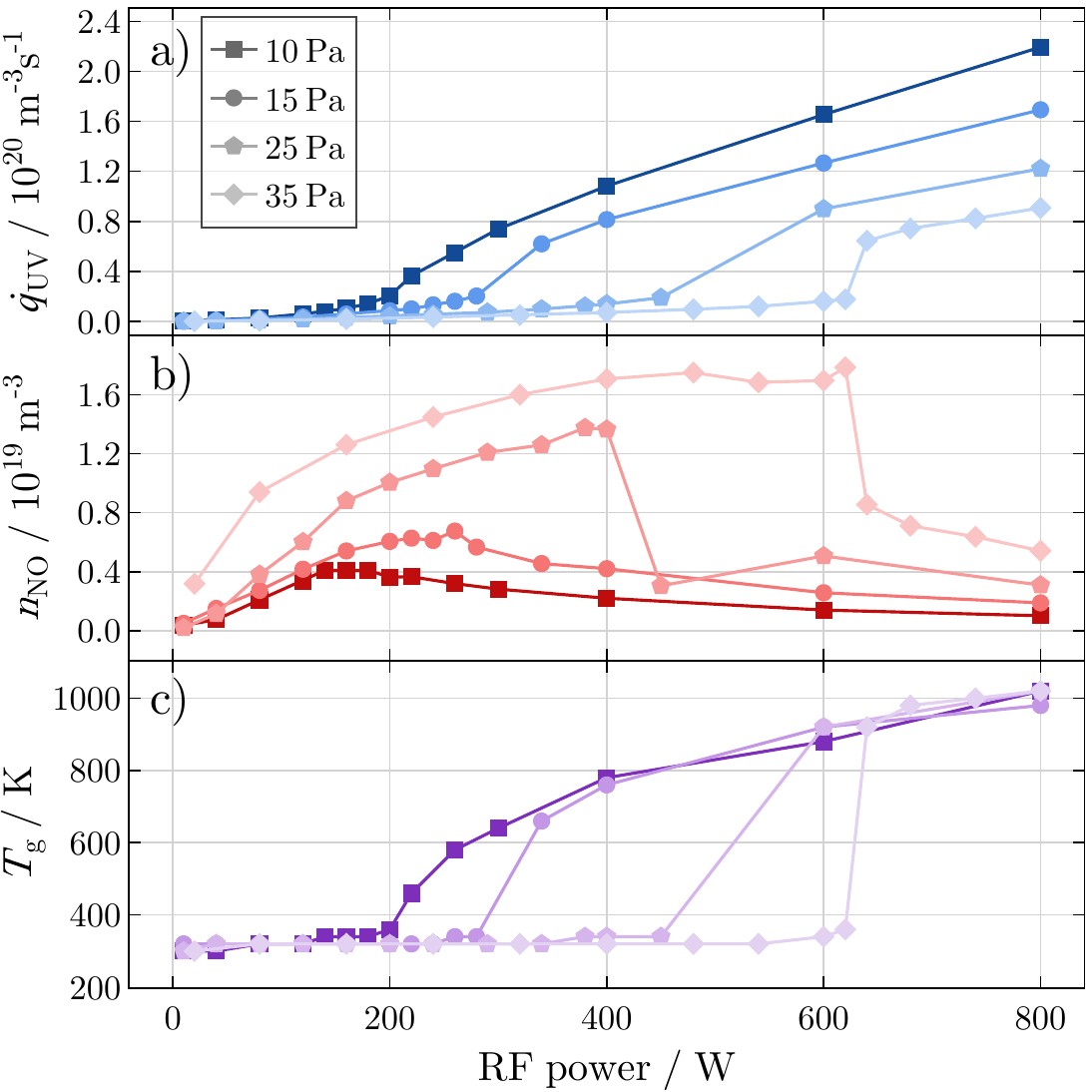}
	\caption{Measured UV photon emission intensity $\dot q_{\mathrm{UV}}$ from 200\,nm to 400\,nm acquired by OES, NO ground state density \textit{n}\textsubscript{NO} acquired by LIF, and neutral gas temperature \textit{T}\textsubscript{g} calculated from OES, over rf power and for a pressure variation of 10\,Pa, 15\,Pa, 25\,Pa, and 35\,Pa. Estimated errors (see chapter \ref{chap:errors}): 12\,\% (UV emission), 21\,\% (NO density), 13.5\,\% (gas temperature).}
    \label{fig:UV_NO_Tg_overPower}
\end{figure}

Figure~\ref{fig:UV_NO_Tg_overPower} shows a) the measured UV photon emission intensity $\dot q_{\mathrm{UV}}$, b) the NO ground state density \textit{n}\textsubscript{NO}, and c) the gas temperature \textit{T}\textsubscript{g}, over a rf power variation from 10\,W to 800\,W and for pressures of 10\,Pa, 15\,Pa, 25\,Pa, and 35\,Pa. Relative uncertainties, determined as described in chapter~\ref{chap:errors}, are given in the figure caption. From the gathered data, it is clear that the E-H mode transition, in the given gas mixture of 16\,sccm (80\,\%) nitrogen and 4\,sccm (20\,\%) oxygen, occurs at around 200\,W for a pressure of 10\,Pa, but only at around 600\,W for a pressure of 35\,Pa. The transition is most easily observed by the point at which the gas temperature begins to increase significantly above room temperature. In general, the gas temperature increases from around 300\,K to 360\,K in E-mode, to up to 1000\,K in H-mode. In addition, the measured parameters exhibit a more gradual response to the transition at lower pressures, while changing abruptly with the transition at higher pressures of 25\,Pa and 35\,Pa. Overall it can be observed that the UV emission increases consistently with the rf power and decreases with the pressure. The emission intensity tends to change relatively continuously over the applied power range, rather than showing an abrupt change at the point of E-H mode transition. \\

The densities of nitric oxide vary between less than $\mathrm{1\,\times\,10^{17}}$\,m\textsuperscript{-3} up to $\mathrm{1.8\,\times\,10^{19}}$\,m\textsuperscript{-3}. While in E-mode, the NO densities increase towards higher powers. However, as soon as the H-mode sets in, the densities abruptly decrease. While in H-mode, the densities decrease with increasing rf power, inversely to the E-mode trends. Despite the decreasing NO densities in H-mode, the UV photon emission, which is attributed to emission from the excited state NO(A), increases consistently. Consequently, the decreased presence of ground state NO(X) in the discharge is likely to be offset by a much higher rate of excitation of the remaining NO, due to the higher electron temperature and density in H-mode. \\

\begin{figure}[t!]
    \centering
    \includegraphics[width=0.495\textwidth]{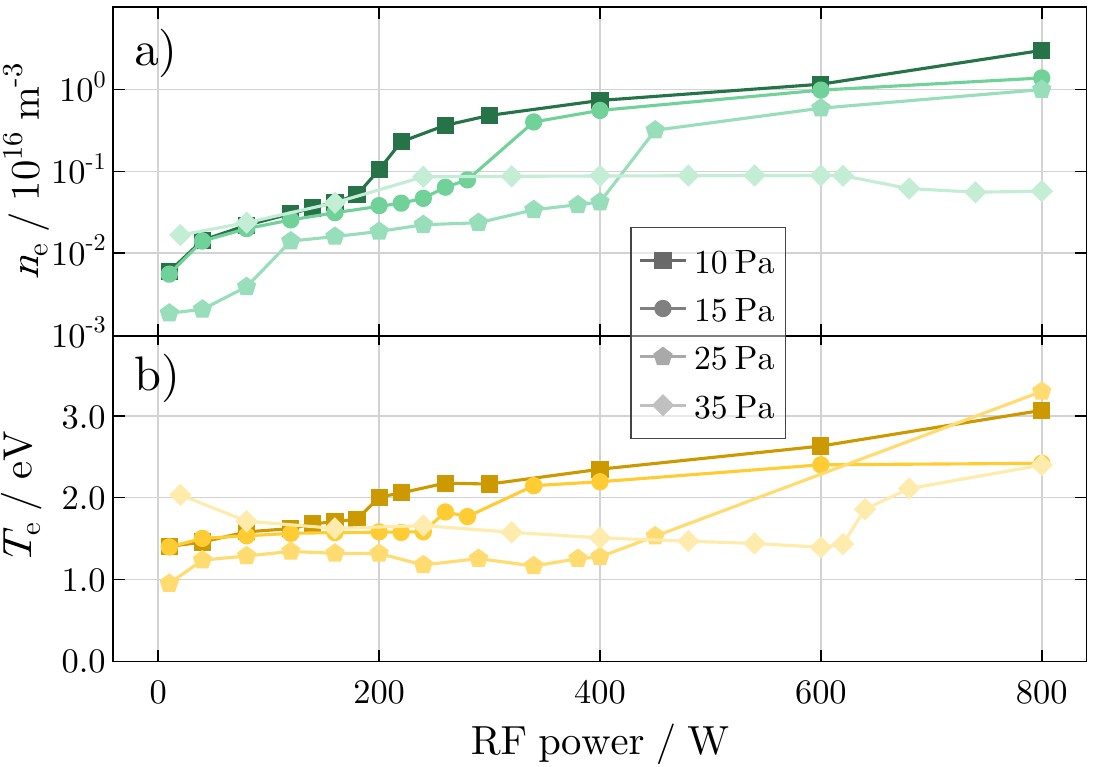}
	\caption{Measured a) electron density \textit{n}\textsubscript{e}, and b) electron temperature \textit{T}\textsubscript{e}, as acquired by MRP over rf power and for a pressure variation of 10\,Pa, 15\,Pa, 25\,Pa, and 35\,Pa. Estimated errors (see chapter \ref{chap:errors}): unknown (electron density and electron temperature).}
    \label{fig:Te_ne_overPower}
\end{figure}

This increase of electron density and electron temperature, as deduced from figure~\ref{fig:UV_NO_Tg_overPower}, is illustrated by the MRP measurements shown in figure~\ref{fig:Te_ne_overPower} a) and b), for the identical parameter range. The E~to~H mode-transition can be observed clearly in the electron density, which increases from less than $\mathrm{1\,\times\,10^{14}}$\,m\textsuperscript{-3} at 10\,W to over $\mathrm{3\,\times\,10^{16}}$\,m\textsuperscript{-3} at a power of 800\,W and a pressure of 10 \,Pa. The increase of the electron density of around one order of magnitude clearly indicates the mode transition. One exception is at a pressure of 35\,Pa, where no increase of the electron density can be observed at the expected power of just over 600\,W. This might be due to limitations within the underlying model of the MRP evaluation method, which was originally developed for idealized electropositive plasmas at pressures in the lower Pascal range. \\

The electron temperature varies more strongly across the tested parameter range, and does not reflect the previously observed trends as strictly. This is most likely due to the method by which the electron temperature is acquired from the electron collision frequency, extracted from the half-width of the measured electron resonance peak by use of the underlying theoretical model. Slight deviations of the computationally determined half-width may lead to significant deviations of the calculated electron temperature, especially when the studied system deviates from those under which the MRP model has originally been developed. Nonetheless, we expect that the general order of magnitude and qualitative trends should hold true, as is also reflected by the qualitative agreement between measurement and simulation, as shown later. Overall, the increasing electron densities and electron temperatures are expected to offset the decreasing NO ground state densities, to result in the continuously increasing NO(A) density and the corresponding UV photon emission intensity, as seen in figure~\ref{fig:UV_NO_Tg_overPower}. \\ 

\begin{figure}[t!]
    \centering
    \includegraphics[width=0.495\textwidth]{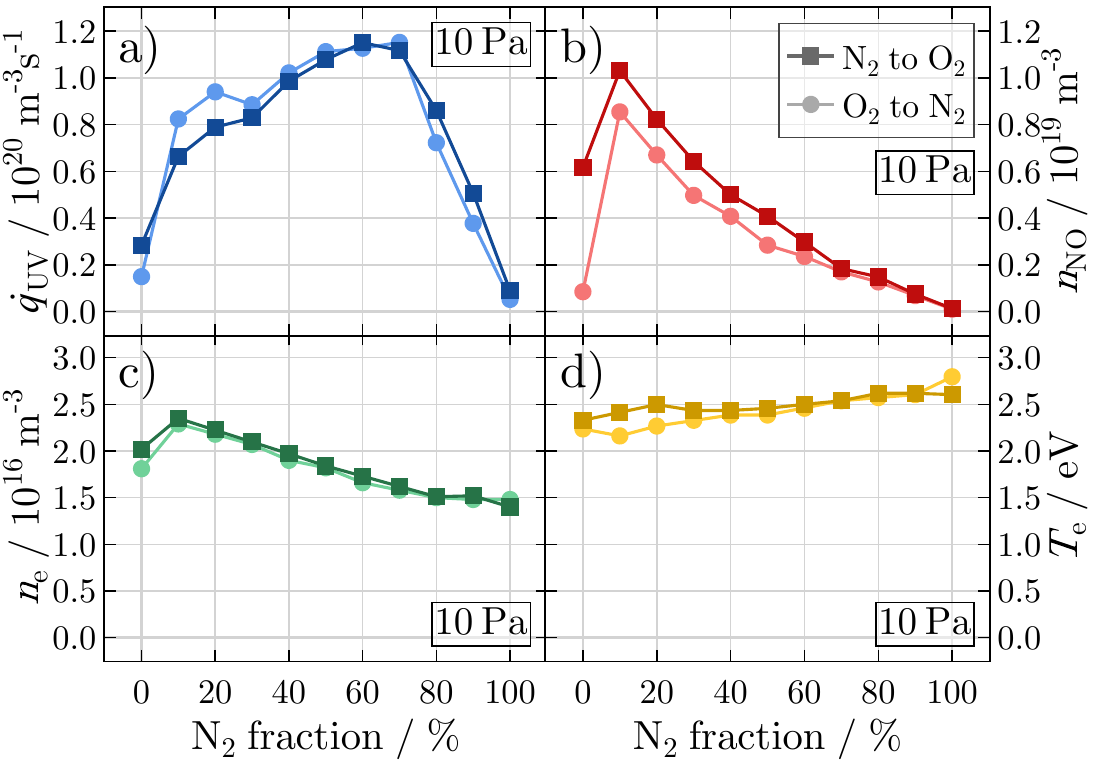}
	\caption{Measured a) UV photon emission intensity $\dot q_{\mathrm{UV}}$ from 200\,nm to 400\,nm acquired by OES, b) NO ground state density \textit{n}\textsubscript{NO} acquired by LIF, as well as c) electron density \textit{n}\textsubscript{e} and d) electron temperature \textit{T}\textsubscript{e} as acquired by MRP for a constant rf power of 500\,W. The gas composition fed into the system is varied from pure nitrogen to pure oxygen (darker traces) and from pure oxygen to pure nitrogen (lighter traces). Estimated errors (see chapter \ref{chap:errors}): 12\,\% (UV emission), 21\,\% (NO density), unknown (electron density and electron temperature).}
    \label{fig:UV_NO_ne_Te_overN2fraction}
\end{figure}

In addition to a variation of power and pressure, a parameter study of the gas composition has been performed. Figure~\ref{fig:UV_NO_ne_Te_overN2fraction} shows the corresponding graphs for a) UV photon emission rate, b) NO density, c) electron density, and d) electron temperature, for a variation of nitrogen to oxygen ratio from 0\,\% to 100\,\%, and vice versa. The pressure was kept at a constant 10\,Pa and the rf power at a constant 500\,W. It is of importance to note that the gas fractions for 100\,\% are expected to be subject to impurities from remaining gas in the chamber, surface coverage, and leak rate, as the gas fraction has been varied over one continuous measurement series, with no intermittent evacuation or cleaning of the chamber. This is also the major reason for why the measurement series have been performed once for 100\,\% of nitrogen to 100\,\% of oxygen, and for 100\,\% of oxygen to 100\,\% of nitrogen, as also indicated by the legend on the figure. \\

Here, the UV photon emission is rather consistent over both variations, exhibiting a plateau at intermediate mixtures of nitrogen and oxygen, with decreasing emission towards pure nitrogen or oxygen. The remaining UV emission in pure N\textsubscript{2} can be attributed to the first negative and second positive system of N\textsubscript{2}, with several emission lines in the range of 300\,nm to 400\,nm. In the case of pure oxygen, however, the remaining emission is caused by the formation of NO from residual N\textsubscript{2} present in the chamber, due to leakage and residual gas. The density measurements of ground state NO in figure~\ref{fig:UV_NO_ne_Te_overN2fraction}\,b) show a distinct maximum at 10\,\% of nitrogen mixed with oxygen, and a consistent decline towards higher nitrogen fractions. A similar quantitative trend for the case of a DC glow discharge was also measured by Gordiets et al. in \cite{gordiets1995}, albeit at higher pressures, and simulated by Guerra et al. in \cite{guerra2001}. Furthermore, the ground state NO density does not decrease to zero for the ``pure" gas mixtures, indicating the importance of even small residual quantities of nitrogen or oxygen, in the respective other gas. Nevertheless, the qualitative trend of the ground state NO density differs from that of the UV photon emission intensity, indicating the importance of electron density and electron temperature, for the corresponding changes in NO(A) excitation rate. Overall, the balance between ground state NO production and NO(A) excitation, leading to UV emission, appears to be rather non-linear, and to depend significantly on the overall plasma chemistry, as dictated by the specific gas mixture and electron properties. \\

The measured electron density, as shown in figure~\ref{fig:UV_NO_ne_Te_overN2fraction}\,c), appears to change relatively little across the entire range of nitrogen to oxygen fraction. A slight decline of the electron density, from a maximum of $\mathrm{2.4\,\times\,10^{16}}$\,m\textsuperscript{-3} at 10\,\% of N\textsubscript{2} to around $\mathrm{1.5\,\times\,10^{16}}$\,m\textsuperscript{-3} at 100\,\% of N\textsubscript{2}, can be observed. The trend is consistent for both variations of the gas composition, i.e. from pure N\textsubscript{2} to pure O\textsubscript{2}, and vice versa. The measured electron temperature increases from around 2.25\,eV in pure O\textsubscript{2} to around 2.75\,eV at in pure N\textsubscript{2}. It should be noted that the electron temperature measured by the MRP, while relatively accurate (see section \ref{chap:errors}), is subject to an unknown error, which likely causes the occasional discrepancies between the two cases of variation of the gas composition. Nevertheless, a consistent trend can be observed. \\

\subsection{Simulation results}
\label{chap:simulation_results}
\begin{figure}[t!]
    \centering
    \includegraphics[width=0.495\textwidth]{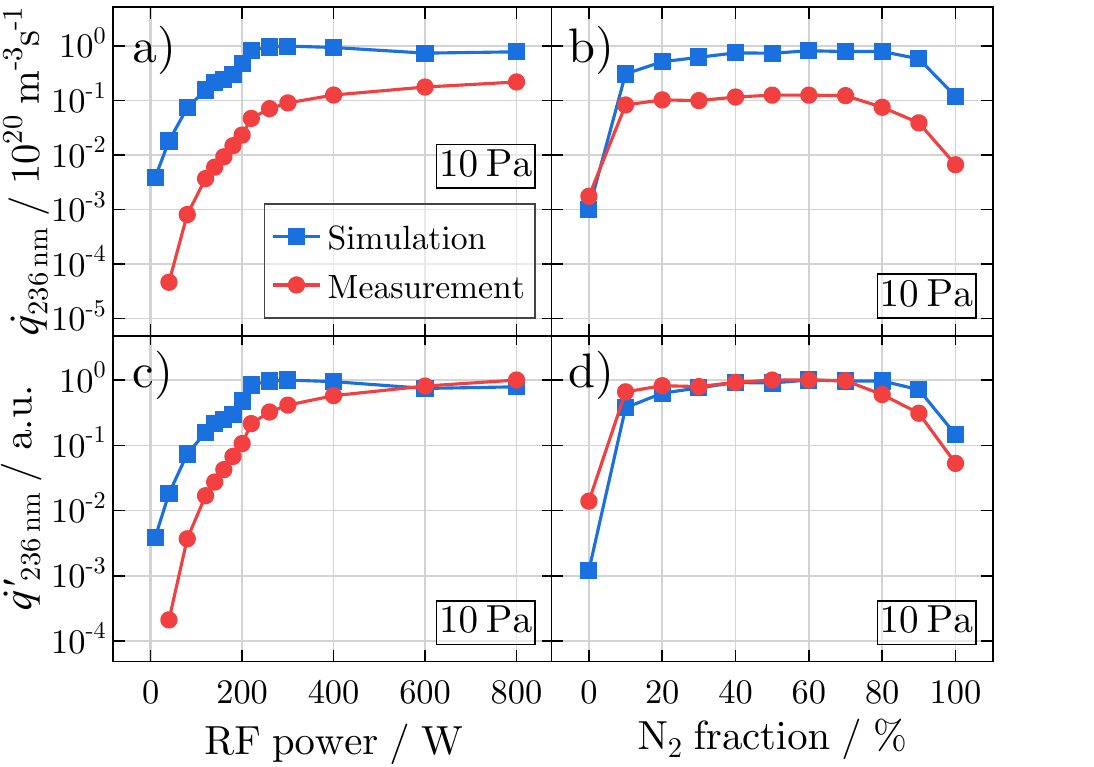}
	\caption{Comparison of measured and simulated UV emission $\dot q_{\mathrm{236\,nm}}$ from NO(A) at the 236\,nm band for a pressure of 10\,Pa over a) rf power and b) gas composition by N\textsubscript{2} fraction. The same traces, individually normalized, are shown as $\dot q'_{\mathrm{236\,nm}}$ in c) and d). The simulated curves are acquired by the solution of the differential equation system given in table\,\ref{tab:reactions}, using \textit{T}\textsubscript{g} as measured by OES, as well as \textit{n}\textsubscript{e} and \textit{T}\textsubscript{e}, measured by the MRP.}
    \label{fig:phot_emiss_simul_overPower_overN2frac}
\end{figure}

Measured and simulated UV photon emission rates of the selected NO(A\textsuperscript{2}\textSigma\textsuperscript{+}, $v=0$\,$\rightarrow$\,X\textsuperscript{2}\textPi, $v=1$) de-excitation band around 236\,nm (integrated from 230\,nm to 237.15\,nm), over rf power and over N\textsubscript{2}:O\textsubscript{2} ratio, are shown in figure~\ref{fig:phot_emiss_simul_overPower_overN2frac} a) and b). Graphs c) and d) of that same figure show identical data, with each trace normalized to its respective maximum, in order to better visualize the qualitative agreement between the traces. Here, the measured gas temperatures, electron temperatures, electron densities, and ground state NO densities given in chapter~\ref{chap:results} are used as input parameters for these calculations. The determined UV photon emission rate of the 236\,nm band, for 10\,Pa and over an rf power from 40\,W to 800\,W, ranges from $\mathrm{5\,\times\,10^{15}}$\,m\textsuperscript{-3}s\textsuperscript{-1} to $\mathrm{2\,\times\,10^{19}}$\,m\textsuperscript{-3}s\textsuperscript{-1}, with a pronounced increase over several orders of magnitude, in E-mode, up until 200\,W and a much less pronounced increase for higher powers, in H-mode. For the variation of gas composition from 100\,\% of oxygen to 100\,\% of nitrogen, as shown in figure~\ref{fig:phot_emiss_simul_overPower_overN2frac}, measurement and simulation agree qualitatively well across the entire range of gas composition, even for the cases of 0\,\% of nitrogen or oxygen, respectively, where no NO and thus no UV emission from the 236\,nm line should be present. As discussed in chapter~\ref{chap:experimental_results}, however, even small impurities in the gas mixture are sufficient to produce measurable densities of ground state NO, which is here used as an input parameter for the simulated excitation and emission processes. This allows for comparison between measurement and simulation for these cases, even in the absence of ground state N\textsubscript{2} or O\textsubscript{2}. \\

\begin{figure}[t!]
    \centering
    \includegraphics[width=0.495\textwidth]{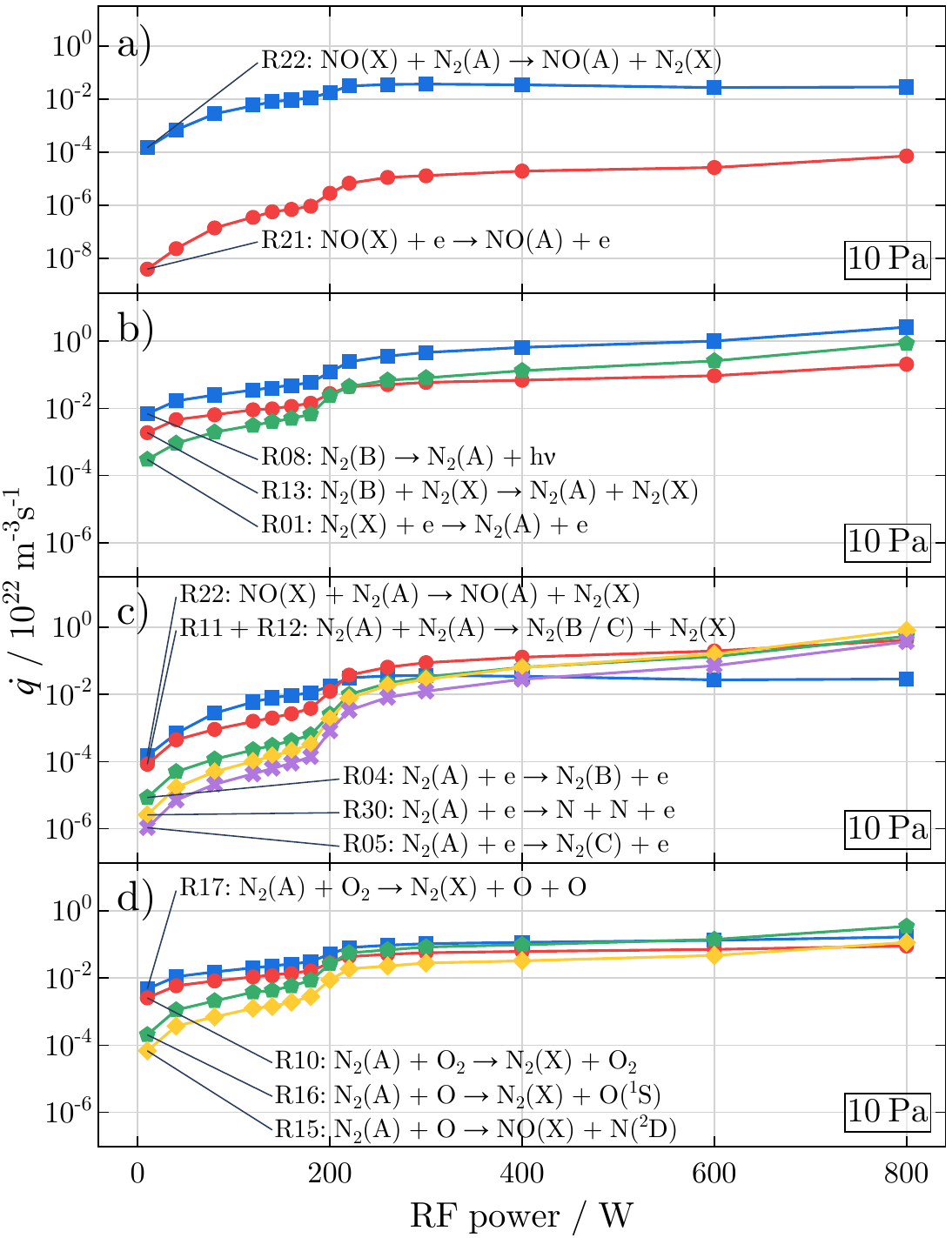}
	\caption{Simulated rates $\dot q$ of the most significant reactions contributing to UV emission from NO(A) at the 236\,nm band. For clarity the reactions are separated into four figures: a) production of NO(A), b) production of N\textsubscript{2}(A), c) consumption of N\textsubscript{2}(A) by collisions with electrons and nitrogen species, and d) losses of N\textsubscript{2}(A) by collisions with oxygen species. The sum of R21 and R22 in figure a) corresponds to the aforementioned total UV emission by NO(A).}
    \label{fig:phot_emiss_simul_react_rates}
\end{figure}

Figure~\ref{fig:phot_emiss_simul_react_rates} shows the rates of the most significant reactions, for the rf power variation at 10\,Pa, which are here used by way of example to explain the qualitative trends observed in figures~\ref{fig:phot_emiss_simul_overPower_overN2frac} a) and c). For clarity the reaction rates shown in figure~\ref{fig:phot_emiss_simul_react_rates} are separated into a) production of NO(A), b) production of N\textsubscript{2}(A), c) consumption of N\textsubscript{2}(A) by collisions with electrons and nitrogen species, and d) losses of N\textsubscript{2}(A) by collisions with oxygen species. Graph a) shows that the main excitation mechanism of NO(X) to NO(A) is by impact with N\textsubscript{2}(A) metastables, rather than by electron impact. Both excitation mechanisms increase in a trend that is qualitatively very similar to the previously discussed photon emission rates, which can likely be attributed to the almost identical trend for the electron density, as shown in figure~\ref{fig:Te_ne_overPower}. With electrons providing the energy to initiate the reaction scheme, this relationship between the electron density and the overall reaction rates is expected, especially with the pronounced transition between E-mode and H-mode. The electron temperature, as shown in figure~\ref{fig:Te_ne_overPower} as well, also increases towards higher rf powers, although not as significantly as the electron density, and is expected to contribute to the discussed trends as well. The decreasing NO(X) density, shown and discussed earlier in figure~\ref{fig:UV_NO_Tg_overPower} and chapter~\ref{chap:experimental_results}, is overcompensated for by the enhanced production of the N\textsubscript{2}(A), as shown in graph b) of figure~\ref{fig:phot_emiss_simul_react_rates}. The loss mechanisms for N\textsubscript{2}(A), which are shown in graphs c) and d) figure~\ref{fig:phot_emiss_simul_react_rates}, generally follow the trend of the increasing N\textsubscript{2}(A) density. However, the relative importance of the different loss mechanisms varies with increasing power. For example, the largest contributions to N\textsubscript{2}(A) consumption at the lowest rf power are collisional quenching with O\textsubscript{2} (R17 and R10). These processes are still important at the highest power, but a number of other reactions also contribute strongly to N\textsubscript{2}(A) consumption, such as electron impact excitation (R04 and R05) and dissociation (R30) as well as metastable-metastable collisions (R11 and R12). \\

While the qualitative agreement between measured and simulated emission rates of figure~\ref{fig:phot_emiss_simul_overPower_overN2frac} is good, there are significant differences in the absolute values between the shown trends, with the simulated values being consistently too high. For the variation of rf power in graph a) the measured and simulated photon emission rates differ by up to two orders of magnitude at the lowest power of 10\,W, and roughly a factor of three at the highest power setting of 800\,W. The discrepancy decreases towards higher powers and H-mode, while in for E-mode the discrepancy between measurement and simulation increases towards lower powers. For the variation of gas composition, there is almost no discrepancy for 0\,\% of N\textsubscript{2}, and an increasing difference to up to one order of magnitude towards 100\,\% of N\textsubscript{2}. There are multiple factors that are expected to contribute to these discrepancies, which come down to either an over-estimation of the N\textsubscript{2}(A) metastable production rate, or an under-estimation of its loss rate. This is due to N\textsubscript{2}(A) being the main species responsible for the excitation of NO(A), and the subsequent spontaneous emission, as shown in \ref{fig:phot_emiss_simul_react_rates}. \\

\begin{figure}[t!]
    \centering
    \includegraphics[width=0.495\textwidth]{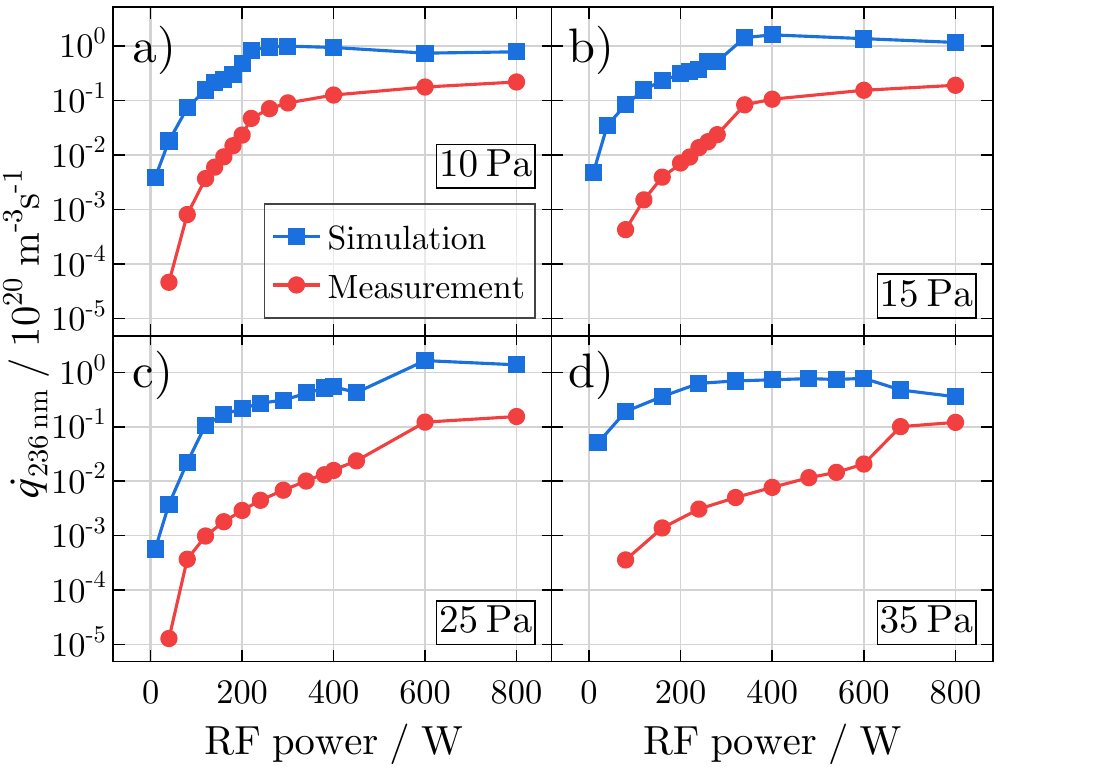}
	\caption{Overview of the measured and simulated UV emission from NO(A) at the 236\,nm band $\dot q_{\mathrm{236\,nm}}$ for a) 10\,Pa, b) 15\,Pa, c) 25\,Pa, and d) 35\,Pa. The simulated curves are acquired by the solution of the differential equation system given in section~\ref{chap:simulation} table~\ref{tab:reactions}, using \textit{T}\textsubscript{g} as calculated from OES, as well as electron density \textit{n}\textsubscript{e} and electron temperature \textit{T}\textsubscript{e}, as measured by MRP.}
    \label{fig:phot_emiss_simul_10_15_25_35Pa}
\end{figure}

In order to gain further insight into the reaction dynamics for such cases, especially with the E-H mode transition shifting over rf power, with changing pressure, the plasma kinetics simulations have been performed for four different pressures of 10\,Pa, 15\,Pa, 25\,Pa, and 35\,Pa, utilizing the experimental results shown in chapter~\ref{chap:experimental_results}. The corresponding graphs are given in figure~\ref{fig:phot_emiss_simul_10_15_25_35Pa}. For all cases, the measured UV emission rates at 236\,nm and the corresponding solutions of the given chemistry set, increase consistently with power. As before, the UV emission rates increase comparably steeply while in E-mode, and flatten down after the transition to H-mode. With increasing pressure, from 10\,Pa to 15\,Pa and 25\,Pa, the measured UV emission rates decrease by less than one order of magnitude. However, the discrepancy between measured and simulated emission rate increases towards higher pressures. This is especially true for the 35\,Pa case, which also exhibits an additional anomaly for the H-mode, at powers higher than 600\,W. Here, the simulated emission rates decrease slightly, while the measured rates increase abruptly. This effect is driven by the lack of an increase in the measured electron densities for H-mode in this case, while the measured electron temperature increases sharply, as shown in chapter~\ref{chap:experimental_results} figure~\ref{fig:Te_ne_overPower}. \\

There are multiple possible reasons for the discrepancy between simulation and measurement, which will be briefly outlined in the following. As discussed in section \ref{chap:errors}, the MRP measured values from the MRP are subject to relatively high uncertainties, especially with respect to the electron temperature, and also increasingly for higher pressures. These discrepancies directly affect the electron properties, which are the main driving mechanism of the chemical kinetics simulation, resulting in corresponding uncertainties. In addition, with respect to the electrons, we assume the EEDF to follow a Maxwellian distribution, which is an approximation that is unlikely to properly capture the distribution in such molecular gas plasmas over a wide range of operating conditions. Test simulations were also performed using a Druyvesteyn distribution function instead. However, this generally led to slightly worse agreement with the measurements. Finally, the reaction scheme itself is subject to a number of major simplifications, which might intensify the deviations between simulations and measurements. Specifically, vibrationally excited N\textsubscript{2} molecules are not included in the reaction scheme, however, these are known to distribute energy and contribute significantly to molecular dissociation in N\textsubscript{2} plasmas. Vibrationally excited N\textsubscript{2} is also of potential importance for the consumption of N\textsubscript{2}(A) \cite{guerra2001}, meaning that neglecting these states could lead to a higher N\textsubscript{2}(A) density, which would be consistent with the overestimation of the UV emission intensity observed in our simulations. Additionally, for practicality, atomic oxygen and the respective excited states are treated in a simplified manner, while the density of atomic nitrogen is not considered in the model, partly due to the fact that vibrationally excited states of nitrogen are known to be important for this, and are themselves not included in the reaction scheme. It is possible that a more detailed inclusion of these factors would lead to a decreased N\textsubscript{2}(A) density, and a better agreement between experimentally measured and simulated UV emission intensity. \\

\section{Conclusion and future work}
\label{chap:conclusion}
A double-inductively coupled plasma device, operated in mixtures of nitrogen and oxygen, is studied with respect to gas temperature, electron density, electron temperature, UV photon emission intensity, and the absolute density of ground state NO. An increase of supplied rf power leads to the expected E-H mode transition, which is shown to have significant impact on the studied parameters. While the gas temperature, ground state NO density, and electron density are observed to change abruptly with the rf power E-H mode transition, the UV photon emission intensity increases more continuously. The UV photon emission is primarily driven by de-excitation of NO(A), which, under the conditions studied in this work, appears to be excited by two mechanisms: electron impact and collisions with nitrogen metastables, in particular N\textsubscript{2}(A). It is shown that with the E-H mode transition the absolute density of ground state NO decreases, yet the UV photon emission increases further, due to higher rates of N\textsubscript{2}(A) formation and changes in reaction pathways compensating for this by increasing the production rate of NO(A). \\

With respect to the gas mixture, given by the ratio of nitrogen and oxygen, the density of ground state NO increases steadily towards lower fractions of nitrogen and higher fractions of oxygen, until approaching a gas mixture composed of almost exclusively oxygen. The UV photon emission shows a different trend, by having a plateau over mixtures from 20\,\% to 80\,\% of nitrogen, and decreasing towards 0\,\% and 100\,\% of nitrogen. This observation affirms the conjecture, that the excitation dynamics of NO are of higher importance to the UV photon emission rate, than the absolute density of the NO ground state. \\

A simple collisional radiative model for the given mixture of molecular nitrogen and oxygen, and a corresponding set of chemical reactions, is solved computationally. The required input parameters are given by the measured densities of nitric oxide and electrons, as well as the electron temperature. At a pressure of 10\,Pa, a typical operating pressure for the studied discharge, a good qualitative match between measurement and simulation is achieved. This holds also true for the variation of gas composition, for nitrogen fractions from 0\,\% to roughly 80\,\%. The reaction rates of the 10\,Pa case reveal that the excitation of NO(X) to NO(A) is primarily driven by collisions with N\textsubscript{2}(A), while the importance of excitation by electron impact increases towards higher rf powers and the correspondingly higher electron densities. Quantitative agreement between measurement and simulation is limited, with the simulation consistently over-estimating the photon emission rates. Agreement improves towards higher rf powers and lower pressures, which is demonstrated by a pressure variation of 10\,Pa, 15\,Pa, 25\,Pa, and 35\,Pa. Possible reasons for this are uncertainties in electron density and temperature, as well as the potentially limited applicability of a Maxwellian EEDF over the range of operating parameters studied. Additionally, vibrational excitation of N\textsubscript{2} was not considered in the chemical kinetics scheme, and treatment of atomic oxygen and nitrogen simplified. \\

Overall it can be concluded that the density of ground state NO, and the UV photon emission rates from the NO(A) state are strongly decoupled in the \mbox{E-H} mode transition in inductively coupled discharges. This decoupling is driven by the balance between gas temperature, electron temperature, and electron density. These parameters affect the excitation and quenching processes of NO(X), NO(A) and N\textsubscript{2}(A) within the discharge. Especially for lower operating pressures, however, the plasma parameters yielded by MRP measurements facilitate a good qualitative match between the measured UV photon emission intensity and the simple collisional radiative model for the UV emission by NO(A). Future studies within this frame could be dedicated to improving the model by adding treatment of vibrationally excited nitrogen states and atomic nitrogen. Furthermore, the chemical kinetics scheme could be coupled with a full global model of the discharge, to further validate the accuracy of the MRP in different operating regimes. This approach could be complemented, for example, by comparison to Langmuir probe or optical emission spectroscopy with a collisional radiative model for molecular nitrogen emission. \\

\section*{Acknowledgement}
This study was funded by the German Research Foundation (Deutsche Forschungsgemeinschaft, DFG) with the Individual Research Grant ``Plasma-inactivation of microbial biofilms" Project number 424927143. The authors would also like to thank House of Plasma GmbH for their continued assistance in use of the multipole resonance probe.

\newcommand{\newblock}{}

\end{document}